\documentclass[webpdf,modern,mediumone]{oup-authoring-template}

\graphicspath{{Fig/}}
\newcommand{\societylogo}{}

\theoremstyle{thmstyleone}%
\theoremstyle{thmstyletwo}%
\theoremstyle{thmstylethree}%

\usepackage{amssymb}
\usepackage{amsmath}
\usepackage{bbm}
\usepackage{pdflscape}
\usepackage{subfiles} 
\usepackage{comment}

\begin{document}

\journaltitle{Journal Title Here}
\DOI{DOI added during production}
\copyrightyear{YEAR}
\pubyear{YEAR}
\vol{XX}
\issue{x}
\access{Published: Date added during production}
\appnotes{Paper}

\firstpage{1}


\title[Short Article Title]{Competing Event Prediction for Amyotrophic Lateral Sclerosis in the Absence of Informative Reference Time}

\author[1,2, $\ast$]{Juliette ORTHOLAND}
\author[1,3]{Stanley DURRLEMAN}
\author[1]{Sophie TEZENAS DU MONTCEL}

\address[1]{\orgdiv{ARAMIS}, \orgname{Sorbonne Universite, Institut du Cerveau - Paris Brain Institute - ICM, CNRS, Inria, Inserm, AP-HP, Hopital de la Pitie Salpetriere}, \orgaddress{\state{Paris}, \country{France}}}
\address[2]{\orgdiv{Medical Informatics Department}, \orgname{Amsterdam UMC, UVA}, \orgaddress{\state{Amsterdam}, \country{Netherlands}}}
\address[3]{\orgname{Qairnel SAS}, \orgaddress{\state{Paris}, \country{France}}}

\corresp[$\ast$]{Corresponding author}

\received{Date}{0}{Year}
\revised{Date}{0}{Year}
\accepted{Date}{0}{Year}





\abstract{Predicting loss of autonomy in chronic disease is critical, but its occurrence may be precluded by death, requiring competing risks methods for accurate prediction. To improve the estimation, joint models incorporate information from longitudinal outcomes when modeling the cumulative incidence function. Their performance, however, typically depends on an informative predefined reference time, often based on disease onset. For many chronic diseases, disease onset is poorly defined or unobserved, limiting the power of existing approaches. \\
We propose a multivariate joint model for longitudinal data and competing risks that accounts for uncertainty in the time origin through a latent disease age and handles potentially high-dimensional longitudinal outcomes through dimension reduction. The model is validated on simulated data and benchmarked against conventional joint models using data from patients with amyotrophic lateral sclerosis, under scenarios with and without age at symptom onset. \\
When symptom onset is unavailable, the proposed model yields substantially improved predictive performance, measured by the integrated Brier score: for initiation of non-invasive ventilation, 0.229(SD 0.044) versus 0.316(SD 0.025)(p = 0.023); for death, 0.165(SD 0.035) versus 0.187(SD 0.042)(p = 0.047). The proposed framework opens new avenues for the dynamic prediction of disease progression when no reliable reference time is available.} 

\keywords{Joint model, Survival, Amyotrophic Lateral Sclerosis, reference time, Longitudinal outcome}





\maketitle


\section{Introduction}


Beyond mortality, other clinically important events can substantially affect patients' lives, including initiation of intensive medication or the occurrence of hazardous events such as stroke. Predicting the occurrence of such events can support timely adaptation of patient management and home-based care even in fatal diseases. Yet, such events may be censored by death, complicating the task of estimating the cumulative incidence function (CIF) of the event of interest.
%
Competing risks models address this issue through two main approaches: cause-specific hazard models, which estimate the CIF via separate cause-specific hazards \citep{prentice_regression_1978}, and the Fine-Gray model, which directly models the subdistribution hazard \citep{fine_proportional_1999}.
To improve the estimation, joint models incorporate information from longitudinal outcomes when modeling the CIF. The two principal joint modeling frameworks are latent class models \citep{proust-lima_joint_2014} and shared random-effects models \citep{rizopoulos_r_2016}, both of which have been extended to competing-risks settings \citep{andrinopoulou_combined_2017, sheikh_joint_2021, li_backward_2023, thomadakis_joint_2023}.

Joint models are typically built upon generalized linear mixed models (GLMMs), which require an informative predefined reference time to align patients' trajectories for estimation \citep{schiratti_bayesian_2017}. Age at symptom onset is often used for this purpose, but may be ill-defined or unavailable in some chronic diseases \citep{young_data-driven_2024}. For example, in amyotrophic lateral sclerosis (ALS), changes in metabolism begin before first motor symptoms \citep{peter_life_2017}, complicating the establishment of reliable reference time points for monitoring disease progression. 
Data-driven progression models address this limitation by estimating latent temporal structures \citep{young_data-driven_2024}: estimating a reference time where GLMMs require an informative predefined one. Although several data-driven models have been proposed for longitudinal data \citep{young_data-driven_2024}, only a few have been developed for time-to-event outcomes \citep{ortholand_joint_2024} and, to our knowledge, none has yet been extended to competing-risks settings.

In this paper, we propose a joint model for competing risks using a data-driven approach. We extend the data-driven joint model of \cite{ortholand_joint_2024} to i) handle multivariate, potentially high-dimensional longitudinal outcomes through dimension reduction, and ii) handle competing events through cause-specific hazards, which give event-specific estimates of association.
In Section~\ref{multi_model}, we describe the proposed model and its estimation procedure. In Section~\ref{sec:simulation}, we validate the model using simulated data by assessing its ability to recover the data-generating process. In Section~\ref{sec:benchmark}, we use real ALS data to compare the model with state-of-the-art ones, with and without an informative predefined reference time for trajectory alignment. 

\section{Method}\label{multi_model}

In this section, we describe how we extend the existing data-driven univariate joint model \citep{ortholand_joint_2024} to a multivariate joint model for competing events. We first introduce the notation, then define the two submodels of the joint model: the longitudinal submodel and the time-to-event submodel, which together form the main contribution of the paper. Finally, we detail the estimation procedure.

\subsection{Notations}\label{multi_notations}

We consider a cohort of $N$ patients, who may experience one of $L$ competing events. The time to the first observed event or censoring is denoted by $t_{e_i}$ and is associated with an event indicator $B_{e_i}$ defined as follows: $B_{e_i} = 0$ if the observation is censored, and $B_{e_i} = l$ if event $l \in \{1, \ldots, L\}$ is observed \citep{andrinopoulou_combined_2017}. In addition, $K$ longitudinal outcomes are measured. For patient $i$, observations are recorded over $n_i$ visits, with $t_{i,j,k}$ denoting the time of visit $j$ for outcome $k$, and $y_{i,j,k}$ the corresponding observed value. 

\subsection{Proposed joint cause-specific model} \label{multi_struct_param}
We detail the main contribution of this paper, which is a joint-modelling approach that does not require an informative predefined reference time. 
The proposed joint cause-specific model builds upon an existing data-driven model for longitudinal data \citep{schiratti_mixed-effects_2015}. We first describe how heterogeneity in longitudinal data is captured by the longitudinal submodel. We then detail how this heterogeneity is incorporated into the time-to-event submodel.

\subsubsection{Longitudinal submodel} \label{model_rm}

For a patient $i$, at visit $j$, we model longitudinal outcome $k$ as a function of time with additive Gaussian noise ($\sigma$):
\begin{align}
     y_{i,j,k} = \gamma_{i,k}(t_{i,j,k}) + \epsilon_{i,j,k}\text{ with }\epsilon_{i,j,k} \sim \mathcal{N} (0, \sigma_k) \label{eq:long_noise}
\end{align}
Longitudinal outcomes in diseases such as ALS often exhibit nonlinear monotonous trajectories with ceiling or floor effects \citep{gordon_progression_2010}. To accommodate these features, we model each longitudinal outcome using a logistic function:
\begin{eqnarray}
      \gamma_{i,k}(t) = \left(1 + \left(\frac{1}{p_k} -1 \right)\cdot \exp\left(-\frac{1}{p_k(1-p_k)} \left ( v_{0,k}(\psi_i(t) - t_0 ) + w_{i,k}\right)\right)\right)^{-1} \label{eq:long}
\end{eqnarray}
where $t_0$ is the population-level estimated reference time, $v_{0,k}$ controls the steepness of the logistic curve at $t_0$, and $p_k$ is the value of the logistic curve at $t_0$. The latent disease age $\psi_i(t)$ aligns the heterogeneous longitudinal outcome and encapsulates the data-driven aspect of this joint model. The spatial shift $w_{i,k}$ controls the relative ordering of the longitudinal outcomes and captures individual-level deviations for each outcome $k$.

\begin{figure}[h]
\begin{center}
    \includegraphics[width=\linewidth]{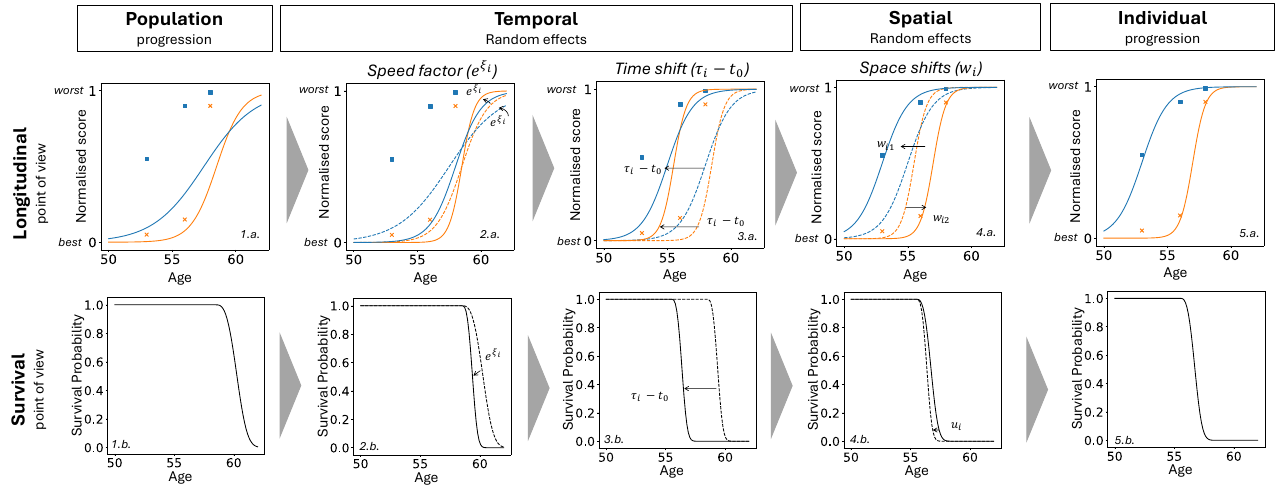}
\end{center}
\caption{Impact of random effects on two longitudinal and one survival outcomes\\ \\
\textit{\underline{Legend:} Panel 2.a. and 2.b.: $e^{\xi_i}$ represents the individual progression speed factor, $e^{\xi_i}>1$: fast progressor, $e^{\xi_i}<1$: slow progressor.\\
Panel 3.a. and 3.b.: $\tau_i$ is the individual-level estimated reference time, $\tau_i - t_0 <0$: early disease onset, $\tau_i - t_0 >0$: late disease onset.\\ 
Panel 4.a. and 4.b.: Longitudinal outcome ordering is captured by latent sources that impact longitudinal outcomes through space-shifts $w_{i,k}$ (panel 4.a.) and time-to-event outcomes through hazard shifts  $u_{i,l}$. $w_{i,k}>0$: early degradation of longitudinal outcome $k$ given a fixed latent age, $w_{i,k}<0$: late degradation of longitudinal outcome $k$ given a fixed latent age; $\exp(u_{i,l})>1$: faster event occurrence, $\exp(u_{i,l})<1$: slower event occurrence.}}
    \label{fig:interpretation}
\end{figure}

\paragraph{Latent disease age - Temporal random effects} The latent disease age anchors patients' progression according to their longitudinal outcome values rather than on time. This data-driven alignment removes the need for an informative predefined reference time \citep{schiratti_mixed-effects_2015}. 
The latent disease age is defined as a function of time $t$:
\begin{align}
    \psi_i(t) = e^{\xi_i}(t - \tau_i) + t_0
\end{align}
where $e^{\xi_i}$ represents the individual progression speed factor (Figure \ref{fig:interpretation} panel 2.a. and 2.b.: $e^{\xi_i}>1$: fast progressor, $e^{\xi_i}<1$: slow progressor), $t_0$ denotes the population-level estimated reference time, $\tau_i$ is the individual-level one. Together, these two estimated reference times align patients on a common disease stage, and their difference can be interpreted as an individual-level temporal shift (Figure \ref{fig:interpretation} panel 3.a. and 3.b.: $\tau_i - t_0 <0$: early disease onset, $\tau_i - t_0 >0$: late disease onset). 

\paragraph{Spatial shifts - Outcome ordering random effects} Patients can exhibit different patterns in the ordering of decline across outcomes. To capture this heterogeneity, each outcome $k$ is associated with a spatial random effect $w_{i,k}$, which shifts the outcome progression along the disease timeline (Figure \ref{fig:interpretation} panel 4.a.: $w_{i,k}>0$: early degradation of outcome $k$ given a fixed latent age, $w_{i,k}<0$: late degradation of outcome $k$ given a fixed latent age). To reduce dimensionality of the longitudinal outcomes and ensure identifiability, we assume spatial shifts arise from a smaller set of $N_s$ latent sources identified using an Independent Component Analysis-like procedure \citep{schiratti_bayesian_2017} (further details in \ref{appendix:ica}). Thus, the sources define the underlying structure, while the spatial shifts provide interpretable summaries of individual-level deviations for each longitudinal outcome. This dimension reduction reduces the number of independent random effects and thus enables modeling of high-dimensional longitudinal outcomes \citep{kaisaridiDeterminingClinicalDisease2025}.

\subsubsection{Time-to-event submodel}\label{multi_model_tte}
The time-to-event submodel captures how variation in the progression of longitudinal outcomes influences the probability and timing of competing events. We model the CIF of event $l$, denoted $F_l(t)$, which represents the probability that event $l$ has occurred by time $t$, in the presence of competing events. Following the cause-specific modeling approach \citep{prentice_regression_1978}, we decompose the CIF in terms of cause-specific hazard functions for each event $j$ ($\lambda_j$), which represent the instantaneous risk of experiencing an event $j$ at time $t$, conditional on having remained event-free up to that time:
\begin{align*}
    F_l(t) = \Pr (t_{e_i}\leq t, J = l) = \int_0^t \lambda_l (u) \exp\left[-\int_0^u \sum_{j=1}^L \lambda_j(s)ds\right]du
\end{align*}
Here, $\exp\left[-\int_0^u \sum \lambda_j(s)ds\right]$ is the probability of remaining event-free up to time $u$. We decompose cause-specific hazards into a baseline hazard reparameterized in terms of the latent disease age, with the longitudinal outcome ordering sources having a proportional effect on the hazard. We model the baseline hazard with a Weibull distribution: 
\begin{align*}
    \lambda_{i,l}(t) =& \lim_{dt \to 0}
\frac{\Pr\!\left(t \leq T_i < t + dt \,\middle|\, T_i \geq t\right)}{dt} \\
=&  \frac{\rho_l \times e^{\xi_i}}{\nu_l}\left( \frac{\psi_i(t)}{\nu_l}\right)^{\rho_l-1} \exp(u_{i,l})
\end{align*}
where $\nu_l$ and $\rho_l$ are the scale and shape parameters of the Weibull; and $u_{i,l}$, the hazard-shift, are a linear combination of the $N_s$ longitudinal outcome-ordering sources $s_{i,m}$:
\begin{align*}
    u_{i,l} = \sum_{m=1}^{N_s} \zeta_{l,m} s_{i,m}
\end{align*}
The hazard shifts capture the effect of the longitudinal outcome-ordering sources on the hazard (Figure \ref{fig:interpretation} panels b.: $\exp(u_{i,l})>1$ faster event occurrence, $\exp(u_{i,l})<1$ slower event occurrence). We introduce sources rather than space-shifts in the survival model to keep the benefit of the longitudinal outcome dimension reduction and potentially deal with high-dimensional longitudinal outcomes. In addition, through this formulation, we assume a shared latent disease age linking the longitudinal and time-to-event processes. 
\subsection{Estimation}

In this section, we detail the estimation procedure for the model parameters and predictions. An implementation of the proposed joint cause-specific model is available in the open-source library Leaspy:  \url{https://github.com/aramis-lab/leaspy}.

\subsubsection{Model estimation algorithm} 

The model is estimated using the Monte Carlo Markov Chain Stochastic Approximation Expectation-Maximization (MCMC-SAEM) algorithm \citep{kuhn_coupling_2004}. This approach overcomes the intractability of the expectation step arising from the model’s nonlinear dependence on random effects. It incorporates a Robbins-Monro stochastic approximation during the final iterations to approximate the posterior means of the parameters \citep{robbins_stochastic_1951}.
MCMC-SAEM convergence is theoretically guaranteed for models belonging to the curved exponential family, to which the proposed model belongs (\ref{suf_stat}). The MCMC-SAEM algorithm is further detailed in \citep{koval_learning_2020} (p.~41–43). The sufficient statistics and update rules required for the implementation of the proposed joint causal-specific model are provided in \ref{suf_stat} and~\ref{max_rule}.

\subsubsection{Likelihood} 

To simplify Equation \ref{eq:long}, we reparametrize the value of the trajectory at population-estimated reference time ($p_k$) by defining $g_k = \frac{1}{p_k}-1$.  All positive parameters are further transformed by a log scale, with $\tilde{x} = \log(x)$. During the stochastic approximation phase of the MCMC-SAEM algorithm, latent parameters ($z$) are drawn from Gaussian distributions:
\begin{itemize}
        \item Latent fixed effects: $z_{fe} = \{ \tilde{g}_k, \tilde{v}_{0,k}, -\tilde{\nu}_l, \tilde{\rho}_l, \beta_{o,m}, \zeta_{l,m}\}$
        \item Latent random effects: $z_{re} = \{ \xi_i, \tau_i, s_{i,m}\}$
    \end{itemize}
To ensure statistical identifiability \citep{lavielle_what_2016}, we set $\overline{\xi} = 0$, $\sigma_{s} = 1$, $\overline{s} =0$ and $t_0 = \overline{\tau}$.
The standard deviations of the latent fixed effects distributions are set to 0.01 by default. These distributions are used to control the exploration of the population parameters during the training phase and do not influence the final parameters \citep{pouletModelingIndividualDisease2023}. The means of the latent fixed effects, together with the means and standard deviations of the random effects and Gaussian noise of longitudinal data, constitute the model parameters:
\begin{align*}
    \theta=\{ \overline{\tilde{g}_k}, \overline{\tilde{v}_{0,k}}, \overline{\tilde{\nu}_l}, \overline{\tilde{\rho}_l}, \overline{\beta_{o,m}}, \overline{\zeta_{l,m}}, \sigma_{\xi}, t_0, \sigma_{\tau}, \sigma \}
\end{align*}
These model parameters are estimated during the likelihood maximization step of the MCMC-SAEM algorithm.
The marginal likelihood is obtained by integrating over the latent parameters: 
\begin{align*}
    p(y,T_e, B_e \mid \theta  ) = \int_{z} p(y,T_e, B_e, z \mid  \theta  ) dz
\end{align*}
We assume conditional independence between the time-to-events and longitudinal processes given the fixed and random effects \citep{rizopoulos_joint_2012}. Under this assumption, the conditional log-likelihood can be decomposed into contributions from the two data processes (time-to-event and longitudinal) and the two types of latent parameters (fixed effects and random effects):
\begin{eqnarray*}
    \log p(y,t_e, B_e, z \mid  \theta  ) =& \log {p(y \mid z, \theta)} + \log p(t_{e}, B_{e} \mid z, \theta) \\
     &+ \log p(z_{re} \mid \theta) + \log p(z_{fe} \mid \theta) 
\end{eqnarray*}
Latent parameters are assumed to follow independent Gaussian distributions. The full expressions for $\log p(z_{re} \mid \theta)$ and $\log p(z_{fe} \mid \theta)$ are provided in ~\ref{annexe_likelihood}.
For the longitudinal process, we assume conditional independence across visits and individuals. Following Equation \ref{eq:long_noise}, the corresponding log-likelihood is:
\begin{align*}
       \nonumber \log {p(y \mid z, \theta, \Pi)} 
       =& \sum_{i,j,k} - \log\left(\sigma_k \sqrt{2\pi}\right) - \frac{1}{2\sigma_k^2} \left( y_{i,j,k} -  \gamma_{i,k}\left( t_{i,j,k}\right)\right)^2
\end{align*}
For the time-to-event process, we also assume independent individuals. The time-to-event log-likelihood accounts for the instantaneous risk of each event and the probability of remaining event-free up to the observed event or censoring time \citep{mozumder_direct_2018}:
\begin{align*}
     \nonumber \log p(t_{e}, B_{e} \mid z, \theta, \Pi) 
    =& &&\sum_{i,l}{\mathbbm{1}_{B_{e_{i}} = l}}  \log \left(\frac{\rho_l e^{\xi_i}}{\nu_l}  \left(\frac{\psi_i(t_{e_i})}{\nu_l}\right)^{\rho_l-1}\exp\left(u_{i,l}\right)\right) \\
    &-&& \sum_{i,l} \left( \frac{\psi_i(t_{e_i})}{\nu_l}\right)^{\rho_l}\exp\left(u_{i,l}\right)
\end{align*}
The likelihood for a single-event setting can be obtained by setting $L=1$. If $\psi_i(t) < t_0$ the corresponding hazard is zero and its logarithm is not finite. To avoid estimation issues, the algorithm is initialized so that $\psi_i(t) > t_0$ for all patients, a strategy inspired by barrier methods \citep{nesterov_lectures_2018}.

\subsubsection{Prediction}

Given the model parameters, the random effects of new patients are estimated by maximizing the log-likelihood with respect to the random effects, excluding the fixed-effect contribution, using the \textit{minimize} solver from the \texttt{Scipy} package  \citep{2020SciPy_NMeth}:
\begin{align*}
    \log {p(y \mid z_{re}, \theta)} + \log p(t_{e}, B_{e} \mid z_{re}, \theta)+ \log p(z_{re} \mid \theta)
\end{align*}
For prediction, the CIF is corrected using the survival probability at the time of the last observation, following common practice \citep{andrinopoulou_combined_2017}.
To accelerate the estimation of the random effects, the Jacobian of the log-likelihood with respect to the random effects can be used. The full expression of the log-likelihood with its Jacobian is provided in \ref{jacobian}.
\section{Simulation Study} \label{sec:simulation}

The simulation study aims to validate the proposed model by i) evaluating the performance of the proposed procedure for selecting the number of latent sources and ii) assessing its ability to recover the data-generating process. It follows the Aims, Data-generating mechanisms, Methods, Estimands, Performance measures (ADEMP) framework \citep{morrisUsingSimulationStudies2019} and builds upon the work of \citep{lavalley-morelle_extending_2024}. 
\subsection{Experimental design}\label{multi_sim_data}

\paragraph{Data generation} Data were generated under the proposed joint cause-specific model, with the first occurring event censoring the other as well as longitudinal outcomes.
We simulated four longitudinal outcomes driven by two sources. The population-estimated reference time and time at baseline were chosen to avoid left-censoring and ensure enough longitudinal data. Other parameters governing disease progression were estimated from the PRO-ACT dataset (described in Section \ref{section_proact}). 
A total of $M=100$ datasets, each comprising $N=300$ patients, were generated. Further details on the data generation process are provided in \ref{sim_method}.
\paragraph{Method} For each simulated dataset, the proposed joint cause-specific model was initialized with parameters from the data-driven longitudinal model fitted for 2,000 iterations. The proposed joint cause-specific model was then fitted for 50,000 iterations (including 10,000 iterations of Robbins-Monro stochastic approximation). We fitted models with 1, 2, and 3 sources. 
\paragraph{Selection of number of sources}
We propose to select the number of sources, driving the dimension reduction of the longitudinal outcomes, using a Bayesian Information Criterion (BIC) adapted for mixed-effects models \citep{delattreNoteBICMixedeffects2014}. To evaluate the method, we computed the BIC for each of them. The model with the higher BIC should be the one with 2 sources, as we generated the data with two sources. 

\paragraph{Ability to recover the data-generating process}
Model performance was assessed based on the accuracy of the estimated model parameters and individual-level random effects. Model parameters were obtained from the Robbins-Monro stochastic approximation.  As the random effects are not subject to Robbins-Monro stochastic approximation, we computed the empirical mean of the last 100 iterations for each subject. \\
Noise in longitudinal outcomes was simulated from a beta distribution, but approximated through estimation with a Gaussian distribution. Consequently, the estimated noise parameter $\sigma$ was compared with the standard deviation of the differences between the simulated observations and the underlying trajectories generated from the simulated random effects. The remaining estimated model parameters and random effects were compared with their corresponding values used for data simulation. 
For each model parameter, Relative Bias (RB), Relative Root Mean Square Error (RRMSE), and Relative Estimation Errors (REE) were computed across the $M$ simulated datasets. Standard errors were assessed using the coverage rate. 
Agreement between simulated and estimated random effects was assessed using the intraclass correlation coefficient (ICC).

\subsection{Results}

Simulated data comprised 300 patients, with 6.9 (SD 3.2) visits per patient over 1.0 (SD 0.5) years. Competing events are observed in 24\% and 9.3\% of the patients, reflecting the characteristics of the real-world data analysed subsequently (Table~\ref{tab:multi_stat_data}). 
The proposed joint cause-specific model required on average 35 minutes per dataset. As expected, the BIC was lower for the model with two sources (-13,368) compared to the one with one (-11,922) and three (-11,499) sources. 

For fixed effects, RB was consistently below 20\% in absolute value, and RRMSE remained below 25\% (Table~\ref{tab:multi_sim_long}). Coverage rates were appropriate given the complexity of the simulated datasets, with 18 of 19 exceeding 80\%, and 8 of 19 exceeding 95\% (Table~\ref{tab:multi_sim_long}). REEs across the 100 datasets were centered around zero (Figure~\ref{fig:multi_ree_sim}). \\
Agreement between the simulated and estimated hazard shifts was limited, with ICCs of 0.479 (0.416) and 0.147 (0.445), likely reflecting the low event rates. For other random effects, the ICC exceeded 0.84 (Table~\ref{tab:multi_simul_intra} in the Appendix).

\section{Benchmark Study}\label{sec:benchmark}

In this section, we illustrate the advantages of the proposed modeling framework, which does not require an informative predefined reference time. To this end, we assess the predictive performance of the proposed joint cause-specific model against that of a shared random effects joint cause-specific model. The comparison is carried out on real ALS data under two scenarios: one in which age at symptom onset serves as the informative predefined reference time, and one in which no informative predefined reference time is available, with inference conducted on the time-since-birth scale.

\subsection{Reference multivariate and cause-specific models}


We benchmarked the proposed joint model against a shared random effects joint model accommodating competing risks, combining a logistic specification for the longitudinal outcomes with a cause-specific hazard model. Estimation was performed using the \texttt{JMbayes2} package in \textsf{R} \citep{rizopoulos_r_2016}. This model is hereafter referred to as the \textit{joint cause-specific JMbayes2 model}. We also compare the longitudinal prediction of the proposed joint model with that of the data-driven longitudinal model described in Section~\ref{model_rm}, fitted using the \texttt{Leaspy} package. The full mathematical specification of the models is provided in Table~\ref{tab:multi_ref_model}.

\subsection{Disease Context and Data}\label{section_proact}

ALS is a neurodegenerative disease from which patients die on average 3 to 4 years after disease onset, primarily due to respiratory failure. Non-invasive ventilation (NIV) is an effective symptomatic therapy \citep{kleopa_bipap_1999}, and represents a major milestone in the patient's loss of autonomy. Here we are interested in NIV initiation prediction, and we considered death as a competing event, and not a semi-competing event, to simplify the fact that NIV initiation further impacts death occurrence. Disease progression is commonly assessed using a clinical measure of functional impairment: the revised ALS Functional Rating Scale (ALSFRS-R) \citep{rooney_what_2017}. Three ALSFRS-R domains were selected to represent distinct aspects of functional decline: bulbar function, fine motor function, and gross motor function \citep{rooney_what_2017}. 

We use data from the 2022 release of the Pooled Resource Open-Access ALS Clinical Trials Consortium (PRO-ACT) database. This database aggregates pseudonymized data from 23 phase II and III clinical trials and one observational study \citep{atassi_pro_act_2014}. Patients were selected based on the availability of age at symptom onset and baseline characteristics (sex, site of disease onset). We further restricted the sample to patients with at least three visits to facilitate the prediction task. To minimize potential bias due to left-censoring of NIV initiation, we included only patients with low baseline impairment, represented by a Mitos score of 0 \citep{fang_comparison_2017}. Two datasets were constructed: one indexed by the age of the patients, without an informative predefined reference time, and the other indexed by disease duration, with age at first symptoms serving as an informative predefined reference time. 

Event times for death, tracheostomy, and NIV initiation were extracted from the PRO-ACT database following the procedure detailed in~\ref{annex_event_extraction}. Death and tracheostomy were treated jointly, as is customary in ALS research. For simplicity, we refer to this composite outcome as death throughout the remainder of this manuscript. 
All longitudinal outcomes were normalized between 0 (representing the healthiest state) and 1 (corresponding to maximum pathological change), using the theoretical minimum and maximum values of each ALSFRS-R subscale \citep{rooney_what_2017}. 

\subsection{Experimental design}

The same benchmarking method was applied to both datasets, with and without informative predefined reference time. A 10-fold cross-validation procedure was applied, with 90\% of the data used for training and 10\% for testing in each fold. For each fold, the proposed joint cause-specific model was initialized with parameters from the data-driven longitudinal model fitted for 2,000 iterations. The proposed joint cause-specific model was then fitted with two sources for 70,000 iterations, including 10,000 iterations of Robbins-Monro stochastic approximation. The joint cause-specific JMbayes2 model was fitted for 25,000 iterations, and the data-driven longitudinal model for 70,000 iterations (including 10,000 iterations of  Robbins-Monro stochastic approximation).

In the test set, random effects were estimated using the first two visits of each patient, and predictions were made for subsequent visits and event times. Longitudinal prediction accuracy was assessed using the mean absolute error (MAE) across the three longitudinal outcomes. 
For the time-to-event outcomes, predictive discrimination was assessed using the concordance index (C-index) at 1 and 1.5 years. Prediction accuracy was additionally assessed using the integrated Brier score (IBS), which evaluates the error in the predicted event probabilities. Time-to-event metrics were computed using the \texttt{sksurv} Python package \citep{sksurv}. Comparisons between models were made using the paired Wilcoxon signed-rank test with Bonferroni correction.
\subsection{Results}

We selected 1,919 patients from the PRO-ACT dataset. A further 42 patients were excluded due to evidence of left-censored NIV initiation. In total, 12,197 longitudinal predictions were made, with predictions occurring at 0.55 years (SD 0.47) after the last observed visit. The proposed joint model required approximately 2 hours per fold, compared with approximately 3.5 hours per fold for the JMbayes2 model. Visualizations of the PRO-ACT data in chronological time and latent disease age are provided in \ref{data_vis}.

For time-to-event predictions using data indexed by age (without informative predefined reference time), the proposed joint cause specific model achieved significantly lower IBS than the joint cause-specific JMbayes2 model for both events (NIV initiation: 0.229 (0.044) vs. 0.316 (0.025) ($p$-value = $2.3 \times 10^{-2}$); death: 0.165 (0.035) vs. 0.187 (0.042) ($p$-value = $4.7 \times 10^{-2}$) (Table~\ref{table:multi_bench_surv}). For data indexed by disease duration (with informative predefined reference time), no significant differences were observed between models (Table~\ref{table:multi_bench_surv}).
\\
For data indexed by age (without informative predefined reference time), the proposed joint cause-specific model achieved significantly lower MAE across all three longitudinal outcomes compared to the data-driven longitudinal model and the joint cause-specific JMbayes2 model.
For data indexed by disease duration (with informative predefined reference time), differences in longitudinal prediction performance were significant only for the gross motor scale. The joint cause-specific JMbayes2 model achieved a significantly lower MAE than the proposed joint cause-specific model (1.369 (1.302) vs. 1.433 (1.351), $p$-value = $2.0 \times 10^{-9}$) (Table~\ref{table:multi_bench_long}).

\section{Discussion}

We introduce the first data-driven multivariate joint cause-specific model for estimating the cumulative incidence function without an informative predefined reference time. By jointly aligning longitudinal and time-to-event processes on a latent disease age, the model captures the temporal structure of disease progression in a fully data-driven manner. In the longitudinal submodel, we summarise the relative ordering of deterioration across longitudinal outcomes with a low-dimensional representation, which is then carried directly into the survival submodel.

On data closely mimicking real ALS trajectories, the extended BIC reliably selected the correct number of latent sources for the dimension reduction of longitudinal outcomes \citep{delattreNoteBICMixedeffects2014}. Fixed-effect parameters were well recovered, though random effects, particularly hazard shifts, were sensitive to data sparsity, reflecting the low event rates inherent to the simulated scenarios.
 
When no predefined reference time was available, the proposed model substantially outperformed the shared random-effects joint model \citep{rizopoulos_r_2016} on both time-to-event calibration (IBS) and longitudinal prediction, demonstrating the concrete advantage of a data-driven alignment strategy. When age at symptom onset was used as reference time, predictive performance was comparable across models, with the shared random-effects model showing a marginal advantage on one longitudinal outcome, likely attributable to its larger parameter space.


The model’s key strength is its ability to align patient trajectories without relying on a potentially unavailable or unreliable disease onset date, a common bottleneck in chronic disease modeling. The shared latent disease age imposes structural assumptions that constrain the parameter space, which is advantageous in rare diseases with many longitudinal outcomes but may limit flexibility in others \citep{young_data-driven_2024}. Future extensions could include spline-based baseline hazards, inclusion of baseline covariates, and accommodation of non-monotonic or interval-censored outcomes. An open-source implementation is available in the Leaspy library \url{https://github.com/aramis-lab/leaspy}.

In summary, the proposed multivariate joint cause-specific model with latent disease age provides a novel framework for modeling disease progression that does not require an informative predefined reference time. This advancement opens new avenues for the dynamic prediction of chronic disease progression.

\section{Conflicts of interest}
The authors declare that they have no competing interests.

\section{Funding}
This research was funded by Agence National de la Recherche (ANR-10-IAIHU-06, ANR-19-P3IA-0001, ANR-19-JPW2-000) and Horizon 2020 Framework Programme (826421).

\section{Data availability}
To contribute to reproducible research the code used is available on the GitLab repository:
\url{https://gitlab.com/JulietteOrtholand/joint_multi} and the model developed was integrated into an
open-source package also available in GitHub: \url{https://github.com/aramis-lab/leaspy}. Finally, data
used in this study came from institutional review board–approved databases, to which patients
consented to participate, as can be requested at \url{https://ncri1.partners.org/PROACT}.

\section{Author contributions statement}

Must include all authors, identified by initials, for example:
J.O., S.D. and S.T. conceived the experiment(s),  J.O. conducted the experiment(s), J.O., S.D. and S.T. analysed the results.  J.O., S.D. and S.T. wrote and reviewed the manuscript.


\begin{landscape}

\begin{table}[h!]
\caption{Specification of benchmarked models\\ \\
\textit{\underline{Legend:} Longitudinal: data-driven longitudinal model, 
joint proposed: the proposed joint cause-specific model, Cause-specific-JMbayes2:  joint model with shared random effects and cause-specific survival model. $spl(t)$: spline function. $L$ events indexed by $l$ and $K$ longitudinal outcomes indexed by $k$. For space reasons, the CIF is not integrated into the table, but all the models followed a cause-specific structure described in \cite{zhang_modeling_2008} and in section \ref{multi_model_tte}}}
\begin{center}
\resizebox{1.3\textwidth}{!}{%
\begin{tabular}{ |c|c|c|c|c|c|c| } 
\hline
Model & Inputs &\multicolumn{2}{|c|}{Effects} & Random effects structure & \multicolumn{2}{|c|}{Link functions}  \\
&  &  Fixed& Random & $\psi_i(t) $ & Longitudinal $\gamma_{i,k}(t)$ & $S_{i,l}(t)$    \\
\hline
Longitudinal  & t & $g, v_0, t_0 $ & $\xi_i, \tau_i, w_i $ & $e^{\xi_i}(t-\tau_i)+t_0$& $\left(1+g_k \times \exp(-{v_{0,k}}\frac{(g_k+1)^2}{g_k}(\psi_i(t)-t_0) + w_{i,k})\right)^{-1}$  & - \\ 


joint proposed & t& $g, v_0, t_0, \rho, \nu$  & $\xi_i, \tau_i, w_i, u_i $ & $e^{\xi_i}(t-\tau_i)+t_0$& $\left(1+g_k \times \exp(-{v_{0,k}}\frac{(g_k+1)^2}{g_k}(\psi_i(t)-t_0) + w_{i,k})\right)^{-1}$  &  $\exp \left(- \left( \frac{\psi_i(t)}{\nu_l}\right)^{\rho_l}\exp\left(u_{i,l}\right)\right)$\\ 

Cause-specific-JMBayes2& t &$\beta_0, \beta_1, g_k, \alpha, spl(t)$& $b_{i,0}, b_{i,1}$& $(\beta_0+b_{i,0}) + (\beta_1+b_{i,1})t$ & $\left(1+g_k \times \exp(\psi_i(t)\right))^{-1}$ &  $\exp \left( -\int_0^t \exp(spl_l(u)+ \sum_k \alpha_k \gamma_{i,k}(u))du\right)$\\ 
\hline
\end{tabular}
}
\end{center}

\label{tab:multi_ref_model}
\end{table}

\end{landscape}

\begin{table}[h!]
\caption{Characteristics of the datasets\\ \\
\textit{\underline{Legend:} Results are presented with mean (SD) [class\%]. \\
Simulated: first dataset simulated over the 100 simulated datasets, PRO-ACT: extraction from PRO-ACT database used to benchmark the models, Lariboisière: extraction from the Lariboisière hospital database used for the application.}}
 \resizebox{\textwidth}{!}{%
\begin{tabular}{|ll|ll|}
\hline
 & &  \multicolumn{2}{c|}{ALS disease}\\
Type & Characteristics                    &          Simulated & PRO-ACT \\
\hline
Number & patients                       &               300 & 1,919\\
 & visits                         &              2,065& 16,036\\
 & patient-years&           287& 1,650\\
 & visits per patients                 &          6.9 (3.2)& 8.4  (4.1)\\
\hline
Time & follow-up (years)                    &          1.0 (0.5)& 0.9 (0.5)\\
 & between visits (months)                  &   2.0 (0.7)& 1.4 (0.7)\\
\hline
Sex                             &  (Male) &   -& 1197 [62.4 \%]\\
Age at baseline                &    &    - & 55.6 (11.4) \\
\hline
Observed events (\%) & Censored event&         72 [24.0 \%]& 477 [24.9\%]\\
 & Death      &          28 [9.3 \%]& 216 [11.3\%]\\
 
\hline
\end{tabular}
}
\label{tab:multi_stat_data}
\end{table}

\begin{table}[h!]
\caption{Validation metrics for the fixed effects of the proposed joint cause-specific model on the Real-like simulated dataset \\ \\
\textit{\underline{Legend:} Simulated: the value of the parameter used for simulation: 4 longitudinal outcomes and 2 competing events, RB(\%): the relative bias, RRMSE (\%): the relative root mean square error, CR(\%): the 95 \% coverage intervals. mean individual log-speed ($\overline{\xi}$) and sources mean and standard deviation ($,\overline{s},\sigma_s$) parameters are not present as they are fixed by the model ($\overline{\xi}=0,\overline{s}=0,\sigma_s=1$) and $t_0 = \overline{\tau}$ } }
\begin{center}
\resizebox{\textwidth}{!}{%
\begin{tabular}{|lllrrrr|}
\hline
\multicolumn{1}{|c}{} & \multicolumn{2}{c}{Parameters name} & \multicolumn{1}{c}{Simulated} & \multicolumn{1}{c}{RB (\%)} & \multicolumn{1}{c}{RRMSE (\%)} & \multicolumn{1}{c|}{CR (\%)}\\ \hline
\multirow{3}{*}{\begin{tabular}[c]{@{}l@{}}Distribution of\\ random effects\end{tabular} }                                        & Estimated reference time (mean)                & $t_0$        &   5.000 &   0.10 &   2.23 &  94.0 [87.4, 97.8] \\
& Estimated reference time (std)           &            $\sigma_{\tau}$               &   1.000 &  -1.47 &   4.40 &  96.0 [90.1, 98.9] \\
& Individual log-speed factor (std)                      & $\sigma_{\xi}$        &   0.790 &  13.21 &  14.85 &  52.0 [41.8, 62.1] \\ \hline
\multirow{12}{*}{\begin{tabular}[c]{@{}l@{}}Longitudinal\\ fixed effects\end{tabular}} & \multirow{4}{*}{Curve values at $t_0$: $\frac{1}{1+g_k}$ ($g_k$)}& $g_0$&  13.958 &  -5.35 &   9.28 &  90.0 [82.4, 95.1]\\
&         & $g_1$&   5.316 &  -8.08 &  11.27 &  81.0 [71.9, 88.2] \\
&     &          $g_2$&   3.993 &  -7.74 &  10.87 &  81.0 [71.9, 88.2] \\
&          & $g_3$&   5.704 &  -5.93 &   7.71 &  74.0 [64.3, 82.3] \\ \cline{2-7} 
& \multirow{4}{*}{Speed of the logistic curves ($v_{0,k}$)}       & $v_{0,0}$&   0.069 &  -7.04 &  10.87 &  84.0 [75.3, 90.6] \\
&                                                     & $v_{0,1}$&   0.188 &  -7.43 &  10.64 &  84.0 [75.3, 90.6] \\
 &         &          $v_{0,2}$&   0.198 &  -8.89 &  11.27 &  78.0 [68.6, 85.7] \\
&              & $v_{0,3}$&   0.112 &  -9.34 &  11.21 &  73.0 [63.2, 81.4] \\ \cline{2-7} 
& \multirow{4}{*}{Estimated noises ($\sigma_k$)}                   & $\sigma_0$&   0.066 &  -3.22 &   4.12 &  78.0 [68.6, 85.7] \\
&                                                     & $\sigma_1$&   0.102 &  -1.08 &   2.44 &  92.0 [84.8, 96.5] \\
&    &          $\sigma_2$&   0.102 &  -0.82 &   2.24 &  94.0 [87.4, 97.8] \\
&      & $\sigma_3$&   0.046 &   0.17 &   2.14 &  95.0 [88.7, 98.4] \\ \hline
\multirow{4}{*}{\begin{tabular}[c]{@{}l@{}}Survival \\ 
fixed effects\end{tabular}}     & \multirow{2}{*}{Weibull scale ($\nu_l$)} & $\nu_0$&   2.800 &  18.04 &  23.19 &  83.0 [74.2, 89.8] \\
&                                                     & $\nu_1$&   3.600 &  -0.09 &   9.97 &  94.0 [87.4, 97.8] \\ \cline{2-7} 
& \multirow{2}{*}{Weibull shape ($\rho_l$)}& $\rho_0$&   1.700 &  -8.38 &  14.15 &  88.0 [80.0, 93.6] \\
&                                                     & $\rho_1$&   2.800 &  10.98 &  19.79 &  92.0 [84.8, 96.5] \\ \cline{2-7} 
\hline
\end{tabular}
}
\end{center}
\label{tab:multi_sim_long}
\end{table}

\clearpage

\begin{table}[h!]
\caption{Prediction performance on time-to-event outcomes using PRO-ACT data\\ \\
\textit{\underline{Legend:} From first symptoms: using the dataset indexed by the years since first symptoms, From birth: using the dataset indexed by the age of the patients, NIV: Non Invasive Ventilation initiation, Joint: the proposed joint model, Longitudinal: data-driven longitudinal model, JMbayes2:  joint model with shared random effects with cause-specific survival model estimated using JMbayes2.\\
Results are presented with the mean (SD) over the 10-fold cross-validation. P-values are computed using a Wilcoxon signed-rank test with Bonferroni correction. \textdownarrow means that the metric should be minimized and \textuparrow maximized. Results in bold are the significantly best for each metric for each event.}}
\begin{center}
\resizebox{\textwidth}{!}{%
\begin{tabular}{|lll|l|ll|}
\hline
 &&& IBS \textdownarrow & C-index 1.0y \textuparrow & C-index 1.5y \textuparrow \\
\hline
\multirow{6}{*}{From first symptoms} &\multirow{3}{*}{NIV}&Joint & 0.219  (0.046) & 0.655  (0.046) & 0.655  (0.046) \\
&&JMbayes2 & 0.226  (0.034) & 0.636  (0.048) & 0.646  (0.049) \\
&&p-value & 1.0e+00 & 2.3e-01 & 1.0e+00 \\
\cline{2-6} 
&\multirow{3}{*}{Death}&Joint & 0.156  (0.032) & 0.656  (0.045) & 0.659  (0.044) \\
&&JMbayes2 & 0.149  (0.02) & 0.636  (0.048) & 0.646  (0.049) \\
&&p-value & 1.0e+00 & 1.6e-01 & 7.7e-01 \\
\hline
\hline

\multirow{6}{*}{From birth} &\multirow{3}{*}{NIV}&Joint & \textbf{0.229  (0.044)} & 0.640  (0.038)& 0.637  (0.036) \\
&&JMbayes2 & 0.316  (0.025) & 0.634  (0.043) & 0.634  (0.043) \\
&&p-value & 2.3e-02 & 1.0e+00 & 1.0e+00 \\
\cline{2-6} 
&\multirow{3}{*}{Death}&Joint & \textbf{0.165  (0.035)} & 0.643  (0.039) & 0.641  (0.037)\\
&&JMbayes2 & 0.187  (0.042) & 0.634  (0.043) & 0.634  (0.043) \\
&&p-value & 4.7e-02 & 1.0e+00 & 1.0e+00 \\
\hline

\end{tabular}
}
\end{center}
\label{table:multi_bench_surv}
\end{table}


\bibliographystyle{oup-abbrvnat}
\bibliography{biblio}

\clearpage

\begin{appendices}

\section{Mathematical appendices}

\subsection{ICA-like dimension reduction}\label{appendix:ica}

To reduce dimensionality, we assume the space shifts arise from a smaller set of latent sources $w_i = A s_i$, where $A$ maps latent sources $s_i$ to outcome-specific shifts $w_i$. The dimension reduction is done using Independent Component Analysis (ICA), which maximizes statistical independence between components. To ensure identifiability, $A$ is constrained to lie in the hyperplane of the mean progression direction $\text{Span}(v_0)$ and the reduce dimensionality can not exceed $(K-1)$ so that $A$ is constructed as a linear combination of the basis vectors $(B_o)_o$ of this hyperplane: $A_m = \sum_{o=1}^{K-1} \beta_{o,m} B_o \quad$ so that $\quad A = (B\beta)^\top$ with $\beta$ a $(K - 1) \times N_s$ matrix of coefficients (\cite{schiratti_bayesian_2017}).

\subsection{Likelihood}\label{annexe_likelihood}

\paragraph{Notations} As a reminder, note that there are $N$ patients indexed by $i$ and each has $n_i$ visits indexed by $j$. Note that $t_0 = \overline{\tau}$ .$K$ outcomes and $N_s$ sources and $T_e$ is the time of observation of the event and $B_e$ is the associated boolean whether the event was observed or not (see section \ref{multi_notations}), parameters are defined in section \ref{multi_struct_param}.

\paragraph{Likelihood} The likelihood estimated by the model is the following:
\begin{align*}
     \nonumber p(y,T_e, B_e \mid \theta, \Pi) =& \int_{z} p(y,T_e, B_e, z \mid  \theta, \Pi) dz
\end{align*}

$p(y, T_e, B_e, z \mid  \theta, \Pi)$ can be divided into four different terms: the longitudinal data attachment, the survival data attachment, and two terms for the prior attachment of latent parameters (fixed and random). We end up with the following expression : 
\begin{eqnarray}
     \nonumber \log p((y,t_e, B_e), z \mid \theta, \Pi) =& \log {p(y \mid z, \theta, \Pi)}  + \log p(t_{e}, B_{e} \mid z, \theta, \Pi) \\
     \nonumber +& \log p(z_{re} \mid z_{fe}, \theta, \Pi) + \log p(z_{fe} \mid \theta, \Pi)
\end{eqnarray}
The different parts of the log-likelihood are described below, associated with their different assumptions, with the priors attached to the latent fixed effect, $p(z_{fe} \mid \theta, \Pi)$, separated for longitudinal and survival effects.

\paragraph*{Longitudinal data attachment} To model the longitudinal process, we assumed that patients and their visits are independent conditionally on random effects and that the noise of the process follows a Gaussian distribution. We thus got (\cite{koval_learning_2020} p.175):
\begin{align*}
       \nonumber \log {p(y \mid z, \theta, \Pi)} =& \sum_{i,j,k} \log p(y_{i,j,k}\mid z,  \theta, \Pi)\\
       \nonumber  =& \sum_{i,j,k} - \log\left(\sigma_k \sqrt{2\pi}\right) - \frac{1}{2\sigma_k^2} \left( y_{i,j,k} -  \gamma_{i,k}\left( t_{i,j,k}\right)\right)^2
\end{align*}

\paragraph*{Time-to-event data attachment } To model the survival process, we assumed that all patients were independent and that the modeling of the survival process depended on whether the event was observed or not. Note that the following equation could be interpreted as follows: the patient must have survived till the time of observation (or censure) and then has an instantaneous risk for the observed events (\cite{mozumder_direct_2018}):
\begin{align*}
     \nonumber \log p(t_{e}, B_{e} \mid z, \theta, \Pi) =& && \sum_{i}\log p(t_{e_i}, B_{e_i} \mid z, \theta, \Pi) \\
     \nonumber =& &&\sum_{i,l} {\mathbbm{1}_{B_{e_{i}} = l}} \times \log \left(h_{i,l}(t_{e_i})\right) + \sum_{i,l}  \log \left(S_{i,l}(t_{e_i})\right)  \\
    \nonumber =& &&\sum_{i,l}{\mathbbm{1}_{B_{e_{i}} = l}} \times \log \left(\frac{\rho_l e^{\xi_i}}{\nu_l}  \left(\frac{e^{\xi_i} (t_{e_i}-\tau_i)}{\nu_l}\right)^{\rho_l-1}\exp\left(u_{i,l}\right)\right)\\
    &-&& \sum_{i,l} \left( \frac{e^{\xi_i}(t_{e_i}-\tau_i)}{\nu_l}\right)^{\rho_l}\exp\left(u_{i,l}\right)
\end{align*}
The likelihood for a simple event could be extracted from the above formula by putting $L=1$. If $\psi_i(t) < t_0$ $\log \left(h_i(t_{e_i})\right)=-\infty$, to prevent estimation issues, we initialized the algorithm at a possible point, getting inspiration from barrier methods (\cite{nesterov_lectures_2018}).

\paragraph*{Latent random effects priors attachment} As patients were supposed to be independent of each other, we supposed that random effects were independent conditionally to $z_{fe}$, $\theta$, and $\Pi$. The regularization term associated, with $\overline{\xi} = 0$, $t_0 = \overline{\tau}$, $\overline{s}=0$ and $\sigma_{s}=1$, is then (\cite{koval_learning_2020} p.175):
\begin{align*}
    \nonumber \log p(z_{re} \mid z_{fe}, \theta, \Pi)
    =& && \sum_i\Big({\log p(\tau_i \mid z_{fe}, \theta, \Pi)}\\
        & && + {\log p( \xi_i \mid z_{fe}, \theta, \Pi)} + \sum_m^{N_s} {\log p(s_{i,m} \mid z_{fe}, \theta, \Pi)}\Big)\\
    \nonumber =& - && {N \log\left( \sigma_{\tau} \sqrt{2\pi} \right) - \frac{1}{2\sigma^2_{\tau}}\sum_i (\tau_i - t_0)^2} \\
    \nonumber & - && {N \log\left( \sigma_{\xi} \sqrt{2\pi} \right) - \frac{1}{2\sigma^2_{\xi}}\sum_i (\xi_i - \overline{\xi})^2}\\
    \nonumber & - && {N N_s \log\left( \sigma_{s} \sqrt{2\pi} \right) - \frac{1}{2\sigma^2_{s}}\sum_i\sum_m^{N_s} (s_{i,m})^2}
\end{align*}

\paragraph*{Latent fixed effects priors longitudinal attachment} Each longitudinal latent fixed effect is independently sampled from a posterior distribution. The regularization term associated is then (\cite{koval_learning_2020} p.175):

\begin{align*}
    \log p(z_{fe} \mid \theta, \Pi)
    \nonumber =& &&\sum_k \left({\log p(\tilde{g}_k \mid \theta, \Pi)} + {\log p(\tilde{v}_{0,k} \mid \theta, \Pi)}\right) \\
    \nonumber & + &&  \sum_{o,m}\log p(\beta_{o,m} \mid \theta, \Pi)\\
    \nonumber=& - && {\sum_k \log\left( \sigma_{\tilde{g}} \sqrt{2\pi} \right) - \frac{1}{2\sigma^2_{\tilde{g}}} \left( \tilde{g}_k - \overline{\tilde{g}}_k \right)^2}\\
   \nonumber & - &&{\sum_k \log\left( \sigma_{\tilde{v}_{0}} \sqrt{2\pi} \right) - \frac{1}{2\sigma^2_{\tilde{v}_{0}}} \left( \tilde{v}_{0,k} - \overline{\tilde{v}}_{0,k} \right)^2}\\
   \nonumber & - &&{(K-1)N_s \log(\sigma_{\beta} \sqrt{2\pi}) - \frac{1}{2\sigma_{\beta}^2}\sum_{o,m}(\beta_{o,m} - \overline{\beta}_{o,m})}
\end{align*}

\paragraph*{Latent fixed effects prior event attachment } Each latent fixed effect is independently sampled from a posterior distribution. The regularization term associated is then:
\begin{align*}
    \log p(\tilde{\nu}, \tilde{\rho}, \zeta \mid \theta, \Pi)
    \nonumber =& &&{\sum_l \log p(\tilde{\nu}_l \mid \theta, \Pi)} + {\log p(\tilde{\rho}_l \mid \theta, \Pi)}\\
    \nonumber & + &&{\sum_{l,m} \log p(\zeta_{l,m} \mid \theta, \Pi)}\\
    \nonumber=& -&&{\sum_l \log\left( \sigma_{\tilde{\nu}_l} \sqrt{2\pi} \right) - \frac{1}{2\sigma^2_{\tilde{\nu}_l}} \left( \tilde{\nu}_l - \overline{\tilde{\nu}_l} \right)^2}\\
    \nonumber & -&&{\sum_l \log\left( \sigma_{\tilde{\rho}_l} \sqrt{2\pi} \right) - \frac{1}{2\sigma^2_{\tilde{\rho}_l}} \left( \tilde{\rho}_l - \overline{\tilde{\rho}_l} \right)^2}\\
    \nonumber & -&&{\sum_{l,m} \log\left( \sigma_{\zeta} \sqrt{2\pi} \right) - \frac{1}{2\sigma^2_{\zeta}} \left( \zeta_{l,m} - \overline{\zeta}_{l,m} \right)^2}
\end{align*}

\paragraph{Total formula}
 
\begin{align*}
     \nonumber \log p((y,T_e, B_e), z,\theta \mid \Pi) 
     &=&& \sum_{i,j,k} - \log\left(\sigma_k \sqrt{2\pi}\right) - \frac{1}{2\sigma_k^2} \left( y_{i,j,k} -  \gamma_{i,k}\left(t_{i,j,k}\right)\right)^2 \\ 
     &+&& \sum_{i,l} {\mathbbm{1}_{B_{e_{i}} = l}} \times \log \left(h_{i,l}(t_{e_i})\right) + \sum_{i,l}  \log \left(S_{i,l}(t_{e_i})\right)  \\
    \nonumber & - && {\sum_k \log\left( \sigma_{\tilde{g}} \sqrt{2\pi} \right) - \frac{1}{2\sigma^2_{\tilde{g}}} \left( \tilde{g}_k - \overline{\tilde{g}}_k \right)^2}\\ 
   \nonumber & - &&{\sum_k \log\left( \sigma_{\tilde{v}_{0}} \sqrt{2\pi} \right) - \frac{1}{2\sigma^2_{\tilde{v}_{0}}} \left( \tilde{v}_{0,k} - \overline{\tilde{v}}_{0,k} \right)^2}\\ 
   \nonumber & -&&{\sum_l \log\left( \sigma_{\tilde{\nu}} \sqrt{2\pi} \right) - \frac{1}{2\sigma^2_{\tilde{\nu}}} \left( \tilde{\nu_l} - \overline{\nu_l} \right)^2}\\
    \nonumber & -&&{\sum_l \log\left( \sigma_{\tilde{\rho}} \sqrt{2\pi} \right) - \frac{1}{2\sigma^2_{\tilde{\rho}}} \left( \tilde{\rho}_l - \overline{\rho}_l \right)^2}\\ 
   \nonumber & - &&{(K-1)N_s \log(\sigma_{\beta} \sqrt{2\pi}) - \frac{1}{2\sigma_{\beta}^2}\sum_{o,m}(\beta_{o,m} - \overline{\beta}_{o,m})}\\ 
   \nonumber & -&&{LN_s\log\left( \sigma_{\zeta} \sqrt{2\pi} \right) - \frac{1}{2\sigma^2_{\zeta}} \sum_{l,m} \left( \tilde{\zeta}_{l,m} - \overline{\zeta}_{l,m} \right)^2} \\ 
    \nonumber &-&&{N \log\left( \sigma_{\tau} \sqrt{2\pi} \right) - \frac{1}{2\sigma^2_{\tau}}\sum_i (\tau_i - \overline{\tau})^2} \\ 
    \nonumber &-&&{N \log\left( \sigma_{\xi} \sqrt{2\pi} \right) - \frac{1}{2\sigma^2_{\xi}}\sum_i (\xi_i - \overline{\xi})^2} \\ 
     \nonumber & - &&{N N_s \log\left( \sigma_{s} \sqrt{2\pi} \right) - \frac{1}{2\sigma^2_{s}}\sum_i\sum_m^{N_s} (s_{i,m} - \overline{s})^2} 
\end{align*}

\subsection{Sufficient statistics}\label{suf_stat}

The convergence of the Monte-Carlo Markov Chain Stochastic Approximation Expectation-Maximization (MCMC-SAEM) algorithm has been proven in \cite{kuhn_coupling_2004} for models that lie in the curved exponential family. For such a family of distributions, the log-likelihood can be written as:
\begin{equation}\label{exp_family}
    \nonumber \log \ p(Y, z, \theta, \Pi) = - \Phi(\theta, \Pi) + \langle S(Y, z), f(\theta, \Pi)\rangle + A(Y,z,\Pi)
\end{equation}
where $\Phi$ and $f$ are smooth functions, and $S$ are called the sufficient statistics. The sufficient statistics are to be understood as a summary of the required information from the latent variables ${z}$ and the observations ${Y}$. Our models fall in such a category, and sufficient statistics are described below. Note that for the joint models, the same kind of decomposition was done by \cite{lavalley-morelle_extending_2024}.

The idea is to rewrite the likelihood in the above form to get sufficient statistics. As a reminder, note that there are $N$ patients indexed by $i$ and each has visits indexed by $j$.

\begin{align*}
        &\log q((y,T_e, B_e),z,\theta \mid \Pi) = \\
        & - \sum_{i,j,k}\ln(\sigma_k \sqrt{2 \pi}) - \langle \underbrace{[\| y_{ij} \|^2]_{ij}}_{S_1(y, z)} \underbrace{- 2 [y_{ij}^T \gamma_i(t_{i,j})]_{ij}}_{S_2(y, z)} + \underbrace{[\| \gamma_i(t_{i,j}) \|^2]_{ij}}_{S_3(y, z)} , \frac{1}{2[\sigma^2_k]_k} \mathbf{1}_{\sum_{i,j,k} 1} \rangle \\
        & + [\sum_{i,l} {\mathbbm{1}_{B_{e_{i}} = l}} \times \log \left(h_{i,l}(t_{e_i})\right) + \sum_{i,l}  \log \left(S_{i,l}(t_{e_i})\right)]_i,  \\
        & - KN \ln(\sigma_{\tilde{v}_0} \sqrt{2 \pi}) - \sum \limits_{k=1}^{K} \frac{1}{2 \sigma_{\tilde{v}_0}^2} \overline{\tilde{v}_0}_k^2 + \langle \underbrace{[\tilde{v}_{0,k}^2]_k}_{S_4(y, z)}, -\frac{1}{2\sigma^2_{\tilde{v}_0}} \mathbf{1}_{K} \rangle + \langle \underbrace{[\tilde{v}_{0,k}]_k}_{S_5(y, z)}, \frac{1}{\sigma_{\tilde{v}_0}^2}  [\overline{\tilde{v}_0}_k]_k \rangle \\
        & - K \ln(\sigma_{\tilde{g}} \sqrt{2 \pi}) - \sum \limits_{k=1}^{K} \frac{1}{2 \sigma_{\tilde{g}}^2} \overline{\tilde{g}}_k^2 + \langle \underbrace{[\tilde{g}_k^2]_k}_{S_6(y, z)}, -\frac{1}{2\sigma^2_{\tilde{g}}} \mathbf{1}_{K} \rangle + \langle \underbrace{[\tilde{g}_k]_k}_{S_7(y, z)}, \frac{1}{\sigma_{\tilde{g}}^2}  [\overline{\tilde{g}}_k]_k \rangle\\
        & - (K-1)N_s \ln(\sigma_\beta \sqrt{2 \pi}) - \sum \limits_{o,m}^{} \frac{1}{2 \sigma_\beta^2} \overline{\beta}_{o,m}^2 \\
        &+ \langle \underbrace{[\beta_{o,m}^2]_{o,m}}_{S_8(y, z)}, - \frac{1}{2 \sigma^2_\beta} \mathbf{1}_{(K-1)N_s} \rangle + \langle \underbrace{[\beta_{o,m}]_{o,m}}_{S_9(y, z)}, \frac{1}{\sigma^2_\beta} [\overline{\beta}_{o,m}]_{o,m} \rangle \\
        & -  \ln(\sigma_{\tilde{\nu}} \sqrt{2 \pi}) - \sum \limits_{l=1}^{L} \frac{1}{2 \sigma_{\tilde{\nu}}^2} \overline{\tilde{\nu}_l}^2 + \langle \underbrace{[\tilde{\nu}_l^2]_l}_{S_{10}(Y, z)}, -\frac{1}{2\sigma^2_{\tilde{\nu}}} \mathbf{1}_{1} \rangle + \langle \underbrace{[\tilde{\nu}_l]_l}_{S_{11}(Y, z)}, \frac{1}{\sigma_{\tilde{\nu}}^2}  [\overline{\tilde{\nu}_l}]_l \rangle \\
        & -  \ln(\sigma_{\tilde{\rho}} \sqrt{2 \pi}) - \sum \limits_{l=1}^{L} \frac{1}{2 \sigma_{\tilde{\rho}}^2} \overline{\tilde{\rho}_l}^2  + \langle \underbrace{[\tilde{\rho}_l^2]_l}_{S_{12}(Y, z)}, -\frac{1}{2\sigma^2_{\tilde{\rho}}} \mathbf{1}_{1} \rangle + \langle \underbrace{[\tilde{\rho}_l]_l}_{S_{13}(Y, z)}, \frac{1}{\sigma_{\tilde{\rho}}^2}  [\overline{\tilde{\rho}_l}]_l \rangle \\
        & - LN_s \ln(\sigma_\zeta \sqrt{2 \pi}) - \sum \limits_{l,m}^{} \frac{1}{2 \sigma_\zeta^2} \overline{\zeta}_{l,m}^2 \\
        &+ \langle \underbrace{[\zeta_{l,m}^2]_{l,m}}_{S_{14}(y, z)}, - \frac{1}{2 \sigma^2_\zeta} \mathbf{1}_{LN_s} \rangle + \langle \underbrace{[\zeta_{l,m}]_{l,m}}_{S_{15}(y, z)}, \frac{1}{\sigma^2_\zeta} [\overline{\zeta}_{l,m}]_{l,m} \rangle \\
        & - {N} \log (\sigma_\tau \sqrt{2 \pi}) - \frac{1}{2\sigma_\tau^2} {N} \overline{\tau}^2  + \langle \underbrace{[\tau_i^2]_i}_{S_{16}(Y, z)} , -\frac{1}{2 \sigma_\tau^2} \mathbf{1}_{N} \rangle  + \langle \underbrace{[\tau_i]_i}_{S_{17}(Y, z)} , \frac{1}{\sigma_\tau^2} \overline{\tau} \mathbf{1}_{N} \rangle \\
                \end{align*}
\begin{align*}
        & - N \log (\sigma_\xi \sqrt{2 \pi}) - \frac{1}{2\sigma_\xi^2} N \overline{\xi}^2 + \langle \underbrace{[\xi_i^2]_i}_{S_{18}(Y, z)} , -\frac{1}{2 \sigma_\xi^2} \mathbf{1}_{N} \rangle  + \langle \underbrace{[\xi_i]_i}_{S_{19}(Y, z)} , \frac{1}{\sigma_\xi^2} \overline{\xi} \mathbf{1}_{N} \rangle \\
        & - N N_s \log (\sigma_s\sqrt{2 \pi}) - N \sum \limits_{m=1}^{N_s} \frac{1}{2 \sigma^2_s} \overline{s_m} \\
        &+ \langle \underbrace{[\tilde{s_{il}}^2]_{il}}_{S_{20}(y, z)} , - \frac{1}{2 \sigma^2_s} \mathbf{1}_{N N_s} \rangle + \sum \limits_{m=1}^{N_s} \langle  \underbrace{[\tilde{s}_{im}]_{i}}_{S_{21}(y, z)} , \frac{1}{\sigma_s^2} [\overline{s}] \rangle 
\end{align*}

\subsection{Maximization update rules}\label{max_rule}

To find the update rule of the different parameters, we need to find the new parameter $\theta$ that maximizes the log-likelihood. As expressions are convex in $\theta$, we can simply derive and look for a critical point. We derive the log-likelihood with respect to each maximized fixed effect. Note that only maximized fixed effects are updated by a maximization rule; other parameters are latent variables that are sampled. $\overline{\xi}$ is first maximized and then set to 0 and $\overline{s} = 0$ and $\sigma_s = 1$. As a reminder, note that there are $N$ patients indexed by $i$ and that each of them has $n_i$ visits indexed by $j$. At iteration $c$, we can use $\tilde{S}^{(c+1)}$ computed with the parameters at iteration $c$ and the formula of $S(Y, z)$ to compute the parameters at iteration $(c + 1)$.

See notation in section \ref{multi_notations} and parameters in section \ref{multi_struct_param}, and sufficient statistics section \ref{suf_stat}
\begin{align*}
    (\sigma^2)^{(c+1)} & \leftarrow \frac{1}{N} [\tilde{S}^{(c+1)}_1 - 2 \tilde{S}^{(c+1)}_2 + \tilde{S}^{(c)}_3]^T \mathbf{1}_1\\
    (\overline{\tilde{v}_0}_{k})^{c+1} & \leftarrow \tilde{S}^{(c+1)}_5\\
    (\overline{\tilde{g}}_k)^{c+1} & \leftarrow \tilde{S}^{(c+1)}_7\\
    (\overline{\beta}_{o,m})^{c+1} & \leftarrow \tilde{S}^{(c+1)}_9\\
    (\overline{\tilde{\nu}}_l)^{(c+1)} & \leftarrow \tilde{S}^{(c+1)}_{11}\\
    (\overline{\tilde{\rho}}_l)^{(c+1)} & \leftarrow \tilde{S}^{(c+1)}_{13}\\
    (\overline{\zeta}_{l,m})^{c+1} & \leftarrow \tilde{S}^{(c+1)}_{15}\\
    (\overline{\tau})^{(c+1)} & \leftarrow
    \frac{1}{N}\tilde{S}^{(c+1)}_{17}\\
    (\sigma^2_{\tau})^{(c+1)} & \leftarrow \frac{1}{N} [\tilde{S}^{(c+1)}_{16} - 2 \overline{\tau} \tilde{S}^{(c+1)}_{17}]^T \mathbf{1}_{N} + \overline{\tau}^2\\
    (\overline{\xi})^{(c+1)} & \leftarrow
    \frac{1}{N}\tilde{S}^{(c+1)}_{19}\\
    (\sigma^2_{\xi})^{(c+1)} & \leftarrow \frac{1}{N} [\tilde{S}^{(c+1)}_{18} - 2 \overline{\xi} \tilde{S}^{(c+1)}_{19}]^T \mathbf{1}_{N} + \overline{\xi}^2
\end{align*}

\subsection{Jacobian Likelihood} \label{jacobian}

To faster personalization, gradients are computed for $\xi_i \times \sigma_{\xi}$ and $\tau_i \times \sigma_{\tau}$. Thus, all the equations must be multiplied by the standard deviation at the end to get the implemented formulas.

\subsubsection{Longitudinal data attachment}

\paragraph*{From likelihood} Longitudinal noise is supposed to follow the Gaussian law, we have to derive the following quantity per patient $i$ and visit $j$ for the outcome $k$:
\begin{eqnarray}
     \log p(y_{i,j,k} \mid z, \theta, \Pi)
    \nonumber =& - \log\left(\sigma_k \sqrt{2\pi}\right) - \frac{1}{2\sigma_k^2} \left( y_{i,j,k} -  \gamma_{i,k}\left( t_{i,j,k}\right)\right)^2
\end{eqnarray}
\paragraph*{Jacobian} Using the known formula of the derivative of the logistic function, we get:
\begin{eqnarray}
    \nonumber C_{i,j,k} =& \frac{(1+g_k)^2}{g_k} \left[ y_{i,j,k} -  \gamma_{i,k}\left( t_{i,j,k}\right)\right][\gamma_{i,k}\left( t_{i,j,k}\right)]\left[1-\gamma_{i,k}\left( t_{i,j,k}\right)\right] \\
    \nonumber  \frac{\partial \log p(y_{i,j,k} \mid z, \theta, \Pi)}{\partial \xi_i} =& \frac{1}{\sigma^2} \left(v_{0,k}\psi_i\left(t_{i,j,k}\right) \times C_{i,j,k}\right) \\
    \nonumber  \frac{\partial \log p(y_{i,j,k} \mid z, \theta, \Pi)}{\partial \tau_i} =& -\frac{1}{\sigma^2}\left(v_{0,k} e^{\xi_i} \times C_{i,j,k}\right)\\
    \nonumber  \frac{\partial \log p(y_{i,j,k} \mid z, \theta, \Pi)}{\partial s_{i,m}} =& -\frac{1}{\sigma^2}\left(A_{k,m}  \times C_{i,j,k}\right)
\end{eqnarray}

\subsubsection{Survival data attachment}

\paragraph*{From likelihood} On the other side, the modeling of the survival process depends on whether the event is observed or not for each patient $i$:
\begin{eqnarray}
     \nonumber \log p(t_{e_i}, B_{e_i} \mid z, \theta, \Pi)
    \nonumber =&\sum_{l}{\mathbbm{1}_{B_{e_{i}} = l}} \times \log \left(\frac{\rho_l e^{\xi_i}}{\nu_l}  \left(\frac{e^{\xi_i} (t_{e_i}-\tau_i)}{\nu_l}\right)^{\rho_l-1}\exp\left(\sum_m \zeta_{l,m}s_{i,m}\right)\right)\\
    \nonumber &- \sum_{l} \left( \frac{e^{\xi_i}(t_{e_i}-\tau_i)}{\nu_l}\right)^{\rho_l}\exp\left(\sum_m \zeta_{l,m}s_{i,m}\right)
\end{eqnarray}
\paragraph*{Jacobian} We thus get:
 \begin{eqnarray}
    \nonumber \frac{\partial \log q(t_{e_{i}}, B_{e_{i}} \mid z, \theta, \Pi)}{\partial \xi_i}
    \nonumber =& \sum_{l} {\mathbbm{1}_{B_{e_{i}} = l}}\rho_l 
    - \rho_l \times \left( \frac{\psi_i\left(t_{e_i}\right)}{\nu_l}\right)^{\rho_l} \exp\left(\sum_m \zeta_{l,m}s_{i,m}\right)\\
    \nonumber =& \sum_{l}{\mathbbm{1}_{B_{e_{i}} = l}}\rho_l 
    + \rho_l \times \log \left(S_{i,l}(t_{e_{i}})\right) \\
    \nonumber \frac{\partial \log q(t_{e_{i}}, B_{e_{i}} \mid z, \theta, \Pi)}{\partial \tau_i} =& \sum_{l} -\frac{(\rho_l - 1)}{(t_{e_{i}}-\tau_i)}{\mathbbm{1}_{B_{e_{i}} = l}}  
    -  \frac{\rho_l e^{\xi_i}}{\nu_l}  \left(\frac{ \psi_i\left(t_{e_i}\right)}{\nu_l}\right)^{\rho_l-1}\exp\left(\sum_m \zeta_{l,m}s_{i,m}\right)\\
    \nonumber =& \sum_{l} -\frac{(\rho_l - 1)}{(t_{e_{i}}-\tau_i)}{\mathbbm{1}_{B_{e_{i}} = l}} -h_{i,l}(t_{e_i})\\
    \nonumber \frac{\partial \log q(t_{e_{i}}, B_{e_{i}} \mid z, \theta, \Pi)}{\partial s_{i,m}} =& \sum_l {\mathbbm{1}_{B_{e_{i}} = l}}\zeta_{l,m}  - \zeta_{l,m} \times \left( \frac{\psi_i\left(t_{e_i}\right)}{\nu_l}\right)^{\rho_l} \exp\left(\sum_m \zeta_{l,m}s_{i,m}\right)\\
    \nonumber =& \sum_{l}{\mathbbm{1}_{B_{e_{i}} = l}}\zeta_{l,m} 
    + \zeta_{l,m} \times \log \left(S_{i,l}(t_{e_{i}})\right) 
\end{eqnarray}

\subsubsection{Random effects regularization}

\paragraph*{From likelihood}

\begin{align*}
    \nonumber \log p(z_{re} \mid \theta, \Pi)
    =& - &&{N \log\left( \sigma_{\tau} \sqrt{2\pi} \right) - \frac{1}{2\sigma^2_{\tau}}\sum_i (\tau_i - \overline{\tau})^2} \\
    \nonumber & - &&{N \log\left( \sigma_{\xi} \sqrt{2\pi} \right) - \frac{1}{2\sigma^2_{\xi}}\sum_i (\xi_i - \overline{\xi})^2}\\
    \nonumber & - &&{N N_s \log\left( \sigma_{s} \sqrt{2\pi} \right) - \frac{1}{2\sigma^2_{s}}\sum_i\sum_m^{N_s} (s_{i,m} - \overline{s})^2}
\end{align*}
 
 \paragraph*{Jacobian} We thus get:
 \begin{eqnarray}
    \nonumber \frac{\partial \log q(\xi_i \mid \theta, \Pi)}{\partial \xi_i}  =& \left(\frac{\xi_i - \overline{\xi}}{\sigma_{\xi}}\right)\\
    \nonumber \frac{\partial \log q(\tau_i \mid \theta, \Pi)}{\partial \tau_i} =& \left(\frac{\tau_i - \overline{\tau}}{\sigma_{\tau}}\right)\\
    \nonumber \frac{\partial \log q(s_{i,m} \mid \theta, \Pi)}{\partial s_{i,m}} =& \left(\frac{s_{i,m} - \overline{s}}{\sigma_{s}}\right)
\end{eqnarray}

\clearpage

\newpage

\section{PROACT Data}

\subsection{Event extraction}\label{annex_event_extraction}

\paragraph{Death} All the deaths were stored in one file. For patients without observed death, we censored their deaths with the date of the last visit. 

\begin{figure}[h]
\begin{center}
    \includegraphics[width=\linewidth]{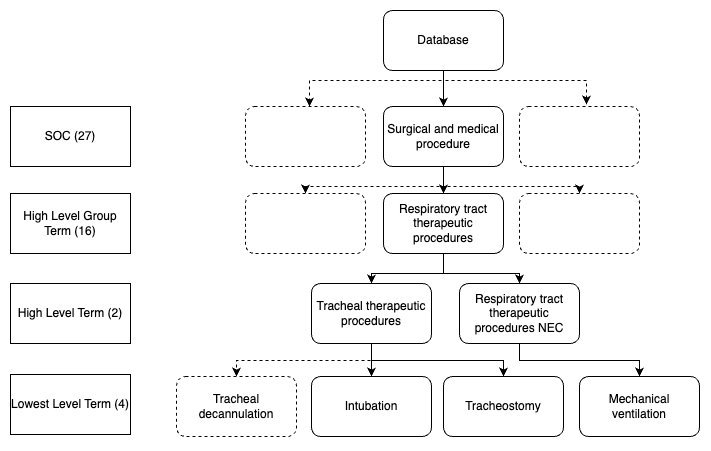}
\end{center}
\caption{Medical intervention terms related to tracheostomy using Medical Dictionary for Regulatory Activities (MDRA)}
\textit{\underline{Legend:} SOC: Standard Of Care, each square represents a level in the medical intervention terms tree, the numbers in parentheses are the numbers of possible terms at each level of the tree, not selected levels and terms are represented with dashed}
    \label{fig:soft_tracheo}
\end{figure}

\paragraph{Tracheostomy} Tracheostomy is a medical intervention; some codes were available in one of the tables using the Medical Dictionary for Regulatory Activities (MDRA). After analysis of the different labels, we have decided to keep the Lowest Level Term, displayed in Figures \ref{fig:soft_tracheo}, to extract an exact date of tracheostomy. Information on tracheostomy is also available in item 12 of ALSFRS-R. For patients with no information from the medical intervention table, we extracted interval-censored events. 

\paragraph{Non-Invasive Ventilation (NIV)} Non-Invasive Ventilation (NIV) was not coded in the medical intervention table but was coded in item 12 of ALSFRS-R and item 10 of ALSFRS. We were thus able to extract interval-censored data. 

\subsection{Visualization}\label{data_vis}

\begin{figure}[!tbp]
  \centering
  \begin{minipage}[t]{0.48\textwidth}
    \includegraphics[width=\textwidth]{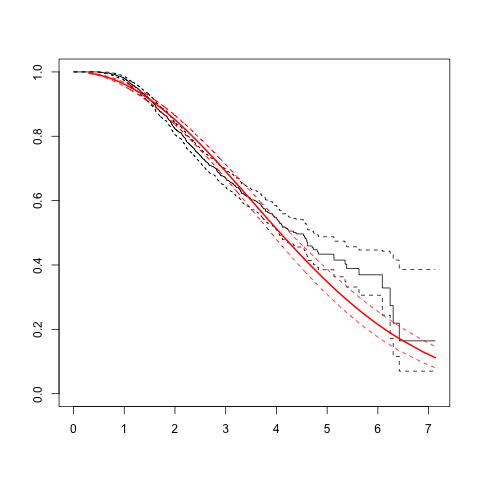}
    \caption{Non-Invasive Ventilation initiation since disease onset from PRO-ACT data}
    \textit{\underline{Legend:} the plot was made using the flexsurvreg package. The black curve represents the event distribution and the red curve a Weibull competing risk model fitted. }
  \end{minipage}
  \hfill
  \begin{minipage}[t]{0.48\textwidth}
    \includegraphics[width=\textwidth]{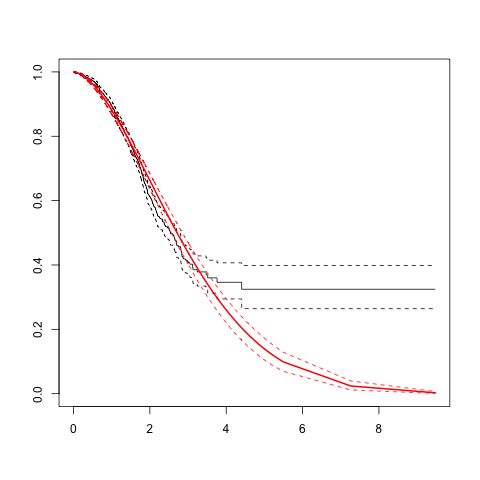}
    \caption{Non-Invasive Ventilation initiation since disease onset in latent disease age, from PRO-ACT data PRO-ACT VNI data reparametrized }
    \textit{\underline{Legend:} the plot was made using the flexsurvreg package. The black curve represents the event distribution and the red curve a Weibull competing risk model fitted. }
  \end{minipage}
\end{figure}

\begin{figure}[!tbp]
  \centering
  \begin{minipage}[t]{0.48\textwidth}
    \includegraphics[width=\textwidth]{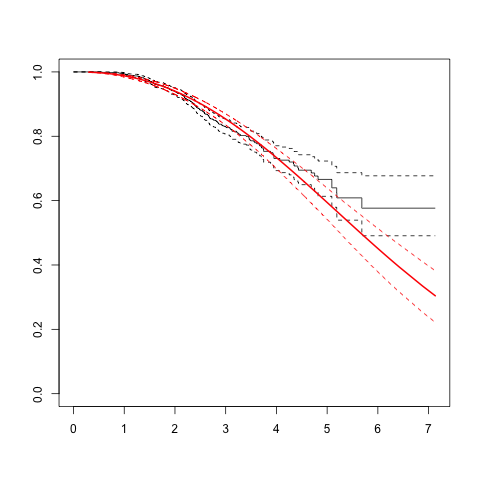}
    \caption{Time to death from disease onset on PRO-ACT data}
    \textit{\underline{Legend:} the plot was made using the flexsurvreg package. The black curve represents the event distribution and the red curve a Weibull competing risk model fitted. }
  \end{minipage}
  \hfill
  \begin{minipage}[t]{0.48\textwidth}
    \includegraphics[width=\textwidth]{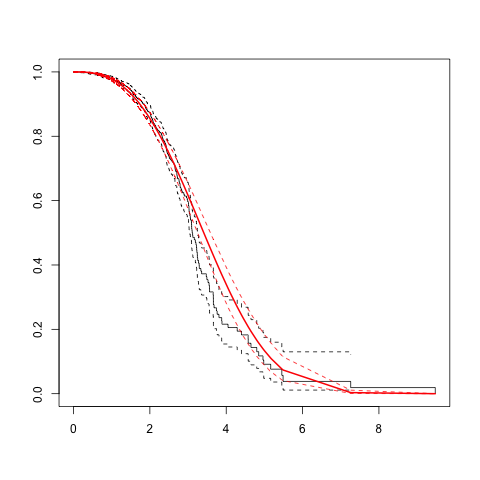}
    \caption{Time to death from disease onset in latent disease age on PRO-ACT data}
    \textit{\underline{Legend:} the plot was made using the flexsurvreg package. The black curve represents the event distribution and the red curve a Weibull competing risk model fitted. }
  \end{minipage}
\end{figure}

\begin{figure}[h]
\begin{center}
    \includegraphics[width=0.8\linewidth]{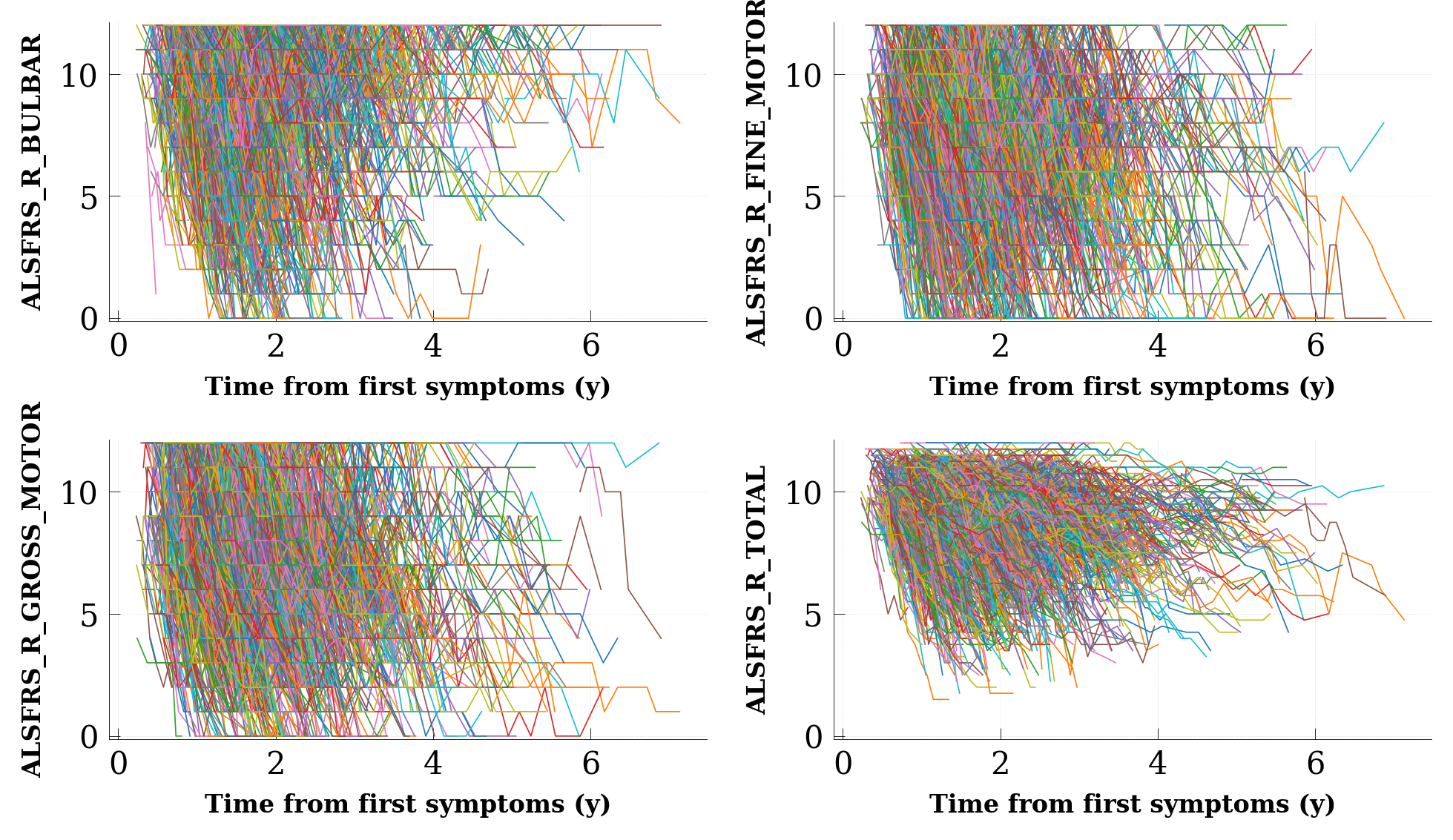}
\end{center}
\caption{PRO-ACT longitudinal data}
\end{figure}

\begin{figure}[h]
\begin{center}
    \includegraphics[width=0.8\linewidth]{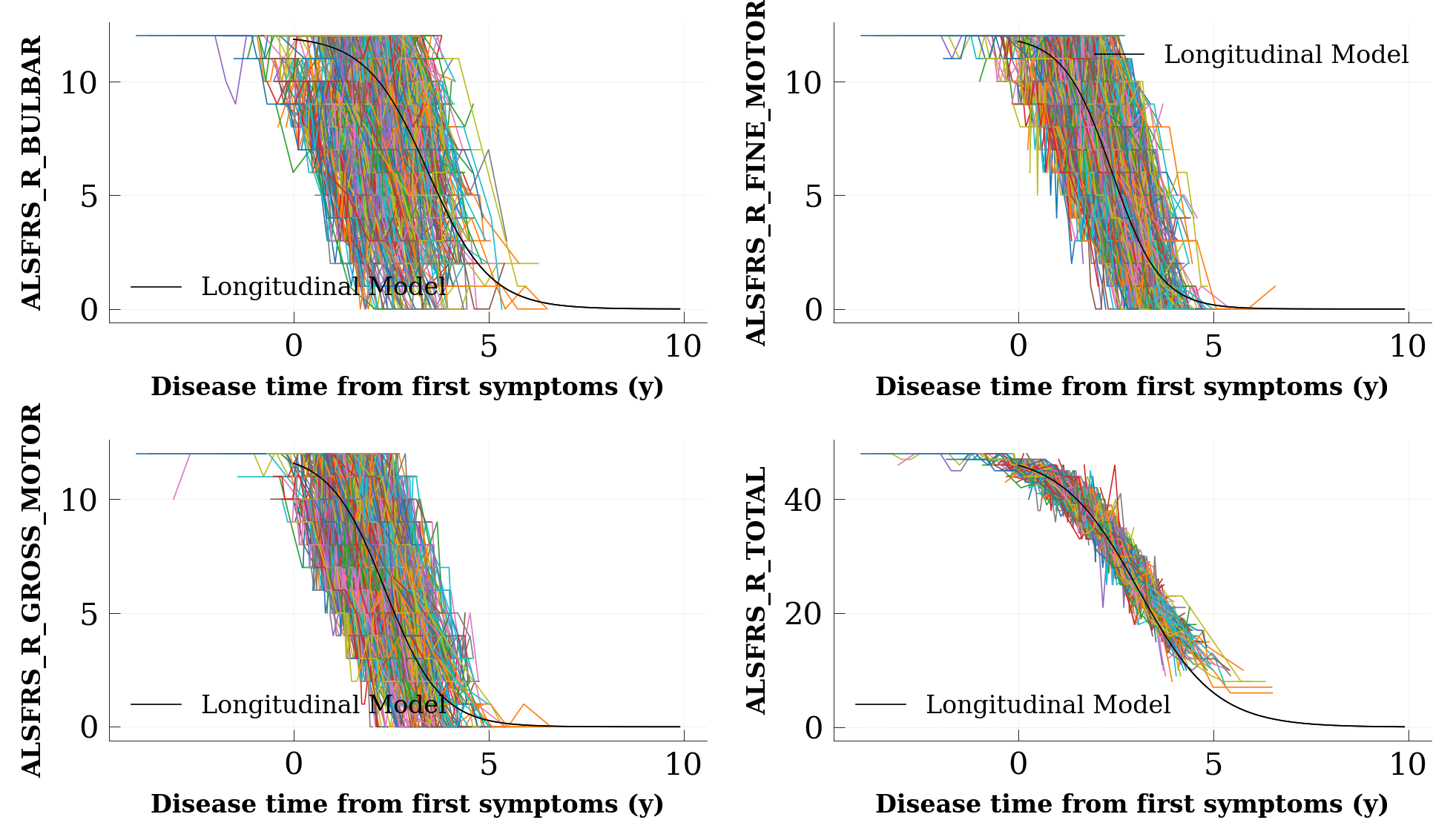}
\end{center}
\caption{PRO-ACT longitudinal data reparametrized in latent disease age}
\end{figure}

\clearpage
\section{Simulation}
\subsection{Method} \label{sim_method}

The simulation procedure comprised the following steps:

\begin{enumerate}
    \item Random effects were drawn as $\xi_i \sim \mathcal{N}(0, \sigma^2_{\xi})$, $\tau_i \sim \mathcal{N}(t_0, \sigma^2_{\tau})$, and $s_{i,m} \sim \mathcal{N}(\overline{s}, \sigma^2_{s})$ with $N_s$ latent sources. 
    \item Baseline age at the first visit $t_{b,i}$ was modelled as $t_{b,i} = \tau_i + \delta_{b_i}$, where $\delta_{b_i} \sim \mathcal{N}(\overline{\delta_b}, \sigma^2_{\delta_b})$.
    \item Follow-up duration per patient $T_{f_i}$ was drawn from $T_{f_i} \sim \mathcal{N}(\overline{T_f}, \sigma^2_{T_f})$. The time interval between two successive visits was drawn as $\delta_{v_{i,j}} \sim \mathcal{N}(\overline{\delta_v}, \sigma^2_{\delta_v})$, and visits were scheduled sequentially until $t_{i,n_i} \leq t_{i,0} + T_{f_i} < t_{i,n_i+1}$.
    \item Longitudinal outcomes were generated at each visit according to $y_{i,j,k} = \gamma_{i,k}(t_{i,j,k}) + \epsilon_{i,j,k}$, where $\epsilon_{i,j,k} \sim \mathcal{B}(0, \sigma^2_k)$. The trajectory function $\gamma_{i,k}(\cdot)$ was defined using the Longitudinal model.
    \item For each patient, cause-specific event times $T_{e_{i,l}}$ were simulated using $T_{e_{i,l}} \sim e^{-\xi_i} \mathcal{W}(\nu_l, \rho_l) + \tau_i$, where $\mathcal{W}(\nu_l, \rho_l)$ denotes a Weibull distribution with scale parameter $\nu_l$ and shape parameter $\rho_l$.
    \item Only the first occurring event for each patient was retained; subsequent events were treated as censored.
    \item Follow-up was terminated at the first observed event: either the end of follow-up or any simulated event. Specifically, measurements with $t_{i,j,k} > T_{e_i}$ were censored, and events occurring after the last visit were censored as well ($\max_j(t_{i,j,k}) < T_{e_i}$).
    \item To ensure that each patient had at least two visits, additional visits were added prior to the first recorded visit or event, at fixed intervals of 1.5 months.
\end{enumerate}

\begin{landscape}

\begin{table}[h!]
\caption{Data simulation parameters by scenario }
\textit{\underline{Legend:} Real-like: Real-like simulated dataset (valid shared latent age hypothesis) (r) indicates when ALS real-like parameters are used. More information on simulation is available in section \ref{multi_sim_data}.$\overline{\tau} = t_0, \overline{\xi} = 0$, $\sigma_{s} = 1$ and $\overline{s} =0$}
\begin{center}
\resizebox{1.3\textwidth}{!}{%
\begin{tabular}{ |c|cc|c|c|c|} 
\hline
 Type & \multicolumn{3}{|c|}{Parameters}   & Estimated & Real-like \\
  & \multicolumn{2}{|c|}{Name} & Symbol & parameters (r)& simulation \\
 \hline
 Patients & {Patient number} &&$N$ & 300 & r\\ 
\hline
\multirow{4}{*}{Random Effect} & {Population estimated reference time (year)}& (mean)&$\overline{\tau}$& 5.00 & r \\ 
& & (std)&$\sigma_{\tau}$ & 1.100& r \\ 
 & \multirow{2}{*}{Individual log-speed factor} &(mean) &$\overline{\xi}$ & 0 & r \\ 
& & (std) &$\sigma_{\xi}$ & 0.790 & r \\
 & Number of sources& & $N_s$ & 2& r\\ 
\hline
Longitudinal & {Speed of the logistic curve} && $v_0$ & [0.069 , 0.188, 0.198, 0.112]  & r \\ 
Fixed Effects & {Curve value at $t_0$: $\frac{1}{1+g}$}&& g & [13.958,  5.316,  3.993,  5.704] & r \\ 
& Estimated noise &&$\sigma$ & [0.07, 0.08, 0.07, 0.04] & [25, 15, 16, 75] \\
& Longitudinal mixing matrix & & $A$ & [[ 0.059, -0.103,  0.001,  0.004], & [[ 0.06, -0.10,  0.00,  0.01],\\
&   & &   &  [ 0.059,  0.006, -0.141, -0.004]] & [ 0.06,  0.006, -0.14, -0.00]]\\
\hline
Survival & {Scale of the Weibull distribution} && $\nu$ & [3.4, 3.9]  & [2.8, 3.6] \\ 
Fixed Effects & {Shape of the Weibull distribution} && $\rho$ & [1.7, 2.8] & r \\
& {Hazard ratio coefficients} && $\zeta$ & [[-0.09, 0.09] & [[-0.09, 0.09]  \\
&  &&   & [-0.1, 0.05]] & [-0.1, 0.0]]\\
\hline
\multirow{5}{*}{Visits}& \multirow{2}{*}{Time between $\tau$ and baseline (years)}& (mean) & $\overline{\delta_{b}}$& 0.2  & 0. \\ 
& & (std) &$\sigma_{\delta_{b}}$& 0.4 & 0.4 \\ 
& \multirow{2}{*}{Time of follow up (year)} &(mean)&$\overline{T_f}$ & 0.8 & 1.1 \\ 
&&(std)&$\sigma_{T_f}$ & 0.5 & r \\ 
& \multirow{2}{*}{Time between visits (months)} &(mean)&$\overline{\delta_{v}}$ & 1.4 & 2.0\\ 
& &(std)&$\sigma_{\delta_{v}}$ & 0.75  & r \\ 
\hline
\end{tabular}
}
\end{center}
\label{table:multi_simulation_param}
\end{table}

\end{landscape}

\subsection{Results}

\begin{table}[h!]
\caption{Intraclass correlation of random effects of the proposed Joint cause-specific model estimated on the Real-like dataset}
\textit{\underline{Legend:} Mean intraclass correlation with the standard deviation over the simulated scenario (SD).}
\begin{center}
\begin{tabular}{|lll|r|}
\hline
   \multicolumn{3}{|c|}{Random effects}&  Intraclass correlation\\
\hline
\multicolumn{2}{|l}{Log-speed factor}&$\xi_i$ &     0.844 (0.020)\\
\multicolumn{2}{|l}{Individual estimated reference time}&$\tau_i$ &     0.919 (0.020)\\
\hline
\multirow{4}{*}{Space shifts}& Bulbar&$w_{i,0}$ &     0.967 (0.006)\\
& Fine motor&$w_{i,1}$ &     0.939 (0.009)\\
& Gross motor&$w_{i,2}$ &     0.954 (0.007)\\
& Total&$w_{i,3}$ &     0.904 (0.042)\\
\hline
\multirow{2}{*}{Survival shifts}& VNI&$u_{i,0}$ &     0.479 (0.416)\\
& Death&$u_{i,1}$ &     0.147 (0.445)\\
\hline
\end{tabular}
\end{center}
\label{tab:multi_simul_intra}
\end{table}

\begin{figure}[h!]
    \begin{center}
    \includegraphics[width=0.5\linewidth]{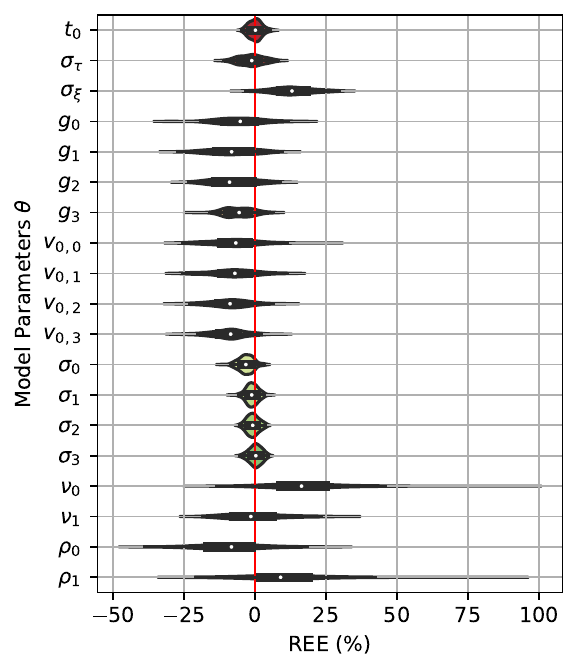}
    \end{center}
    \caption{Relative Estimation Error on the simulated real-like datasets} \label{fig:multi_ree_sim}
    \textit{\underline{Legend:} Distribution over the 100 real-like simulated datasets, $t_0$: Estimated reference time (mean) , $\sigma_{\tau}$: Estimated reference time (std), $\sigma_{\xi}$: Individual log-speed factor (std), $g_k$: Curve values at $t_0$ $\left(\frac{1}{1+g_k}\right)$, $v_{0,k}$: Speed of the logistic curves, $\sigma_k$: Estimated noises, $\nu_l$: Weibull scale, $\rho_l$: Weibull shape. Note that: $\overline{\tau} = t_0, \overline{\xi} = 0$, $\sigma_{s} = 1$ and $\overline{s} =0$}
\end{figure}

\clearpage

\section{Benchmark}

\begin{table}[h!]
\caption{Absolute error on the longitudinal outcomes for the proposed Joint cause-specific and reference models on PRO-ACT data (Benchmark dataset)}
\textit{\underline{Legend:} From first symptoms: using the dataset indexed by the years since first symptoms, From birth: using the dataset indexed by the age of the patients, Joint: the proposed Joint cause-specific model, Longitudinal: data-driven longitudinal model, JMbayes2:  joint model with shared random effects with cause-specific survival model estimated using JMbayes2.\\
Results are presented with the mean (SD) over the 10-fold cross-validation. P-values are computed using a Wilcoxon signed-rank test with Bonferroni correction between the Joint Temporal model and each of the reference models. The absolute bias should be minimized, and the significantly best results compared to the proposed joint cause-specific model are in bold. 12,197 longitudinal predictions were made at 0.55 (0.47) years from the last visit.}
\begin{center}
\resizebox{\textwidth}{!}{%
\begin{tabular}{|ll|r|rr|rr|}
\hline
{} &&   \multicolumn{1}{|c|}{Joint} & \multicolumn{1}{|c}{Longitudinal}  & \multicolumn{1}{c|}{ p-value} & \multicolumn{1}{|c}{JMbayes2} & \multicolumn{1}{c|}{p-value} \\
\hline
\multirow{3}{*}{From first symptoms}&Bulbar       &  1.200 (1.33) & 1.196 (1.324) & 1.0e+00 & 1.18 (1.239) & 7.3e-01 \\
&Fine motor   &   1.514 (1.432) & 1.507 (1.435) & 2.5e-19 & 1.503 (1.395) & 1.0e+00 \\
&Gross motor & 1.433 (1.351) & 1.425 (1.358) & 4.7e-29 & \textbf{1.369 (1.302)} & 2.0e-09\\
\hline
\hline
\multirow{3}{*}{From birth}&Bulbar       &  \textbf{1.196 (1.405)} & 1.218 (1.466) & 1.8e-14 & 1.271 (1.48) & 9.5e-49 \\
&Fine motor   &   \textbf{1.533 (1.512)} & 1.595 (1.639) & 1.0e-16 & 2.003 (1.994) & 2.8e-162 \\
&Gross motor & \textbf{1.435 (1.43)} & 1.482 (1.528) & 3.1e-19 & 1.85 (1.783) & 1.7e-170 \\
\hline
\end{tabular}
}
\label{table:multi_bench_long}
\end{center}
\end{table}

\clearpage

\end{appendices}

\end{document}